# Space-Filling Fractal Description of Ion-induced Local Thermal Spikes in Molecular Solid of ZnO

*Shoaib Ahmad[1,2,*] and Muhammad Yousuf[1]*

[1]CASP, Church Road, Government College University, Lahore 54000, Pakistan

[2]National Center for Physics, QAU Campus, Shahdara Valley, Islamabad 44000, Pakistan

*Corresponding Author: sahmad.ncp@gmail.com

ABSTRACT: Anions of the molecules ZnO, $O_2$ and atomic Zn and O constitute mass spectra of the species sputtered from pellets of molecular solid of ZnO under $Cs^+$ irradiation. Their normalized yields are independent of energy of the irradiating $Cs^+$. Collision cascades cannot explain the simultaneous sputtering of atoms and molecules. We propose that the origin of the molecular sublimation, dissociation and subsequent emission is the result of localized thermal spikes induced by individual $Cs^+$ ions. The fractal dimension of binary collision cascades of atomic recoils in the irradiated ZnO solid increases with reduction in the energy of recoils. Upon reaching the collision diameters of atomic dimensions, the space-filling fractal-like transition occurs where cascades transform into thermal spikes. These localized thermal spikes induce sublimation, dissociation and sputtering from the region. The calculated rates of the subliming and dissociating species due to localized thermal spikes agree well with the experimental results.

**TOC GRAPHICS**

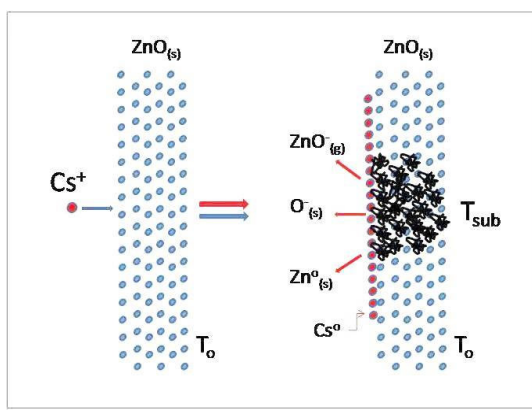

We report results of mass spectroscopy of sputtered anions from $Cs^+$-irradiated ZnO solid that include atomic as well as molecular species. Sputtering of the atomic species (Zn and O) is normally assumed to be the result of binary collision cascades intersecting the upper outer surface but the emission of ZnO requires sublimation. Binary collision-based sputtering theories[1-3] predict energy dissipation by the recoils to occur in a region of the order of the irradiating ion's range. The evolution of cascades is not expected, in these theories, to raise the local temperature. Typical energies of sputtered atoms are of the order of targets' surface binding energies[2]. Thermal spikes on the other hand, have been proposed to describe the emission of atoms from irradiated metals with thermal energies[4-10]. Sputtering of a molecular solid like ZnO by thermal spikes presents the dual route for the sublimation and dissociation that lead to the emission of ZnO, O and Zn. Fractal-geometry approach for the transition from binary collision cascades to thermal spikes has shown[11] that the fractal dimension of cascade increases as a result of the change of effective interatomic potential. Thermal spikes were proposed to arise when the fractal dimension increases and becomes space-filling. This happens when each participating particle in the spike region has kinetic energy of few eV. We recently reported that carbon cluster emissions from single walled carbon nanotubes may be the result of ion-induced localized thermal spikes[12]. The conclusions regarding the heavy ion-induced local thermal spike were based on the calculations of the probabilities of the clusters subliming from nanotubes. However, we did not propose the oriin of the thermal spikes. We show in this communication that one can justify the fractal description of the space-filling binary collision cascades transforming into localized thermal spikes will lead to the sublimation and dissociation of ZnO.

Figure 1 shows the mass spectra of the sputtered species for the range of $Cs^+$ energies $E(Cs^+)$ from 1.0 to 5.0 keV. The species with dominant intensities are $O^-$ and $ZnO^-$ with $O_2^-$ and $Zn^-$ as the minor species. The SNICS output are anions—a fact that explains the relatively small number densities of $Zn^-$. ZnO solid dissociates as $ZnO_s \rightarrow ZnO_g \rightarrow Zn_g^{++} + O_g^{--}$  (1). This reaction implies the need for multiple electron capture processes to take place before Zn can escape the surface as anion. While $O^-$ will naturally emerge as a negative ion. That explains the preponderance of $O^-$ and the diminished intensities of $Zn^-$ as shown in the mass spectra of Figure 1. Only in the last spectrum we have plotted the anionic intensities on log scale to highlight the presence of $Zn^-$ peak with two orders of magnitude lower intensity.

In the inset of Figure 1, the normalized yields $N_x/\Sigma N_x$ are plotted as a function of E(Cs$^+$) where $x$ stands for O, Zn, O$_2$ and ZnO. The normalized yields of all the sputtered species in the figure are insensitive to the variations in the energy of the irradiating Cs$^+$. This aspect of the experimental results is essential to the understanding and interpretation of the atomic and molecular emissions from ZnO solid. The linear collision cascades have a direct dependence on irradiating ion energy[1-3], which has not been seen to be the case in Figure 1.

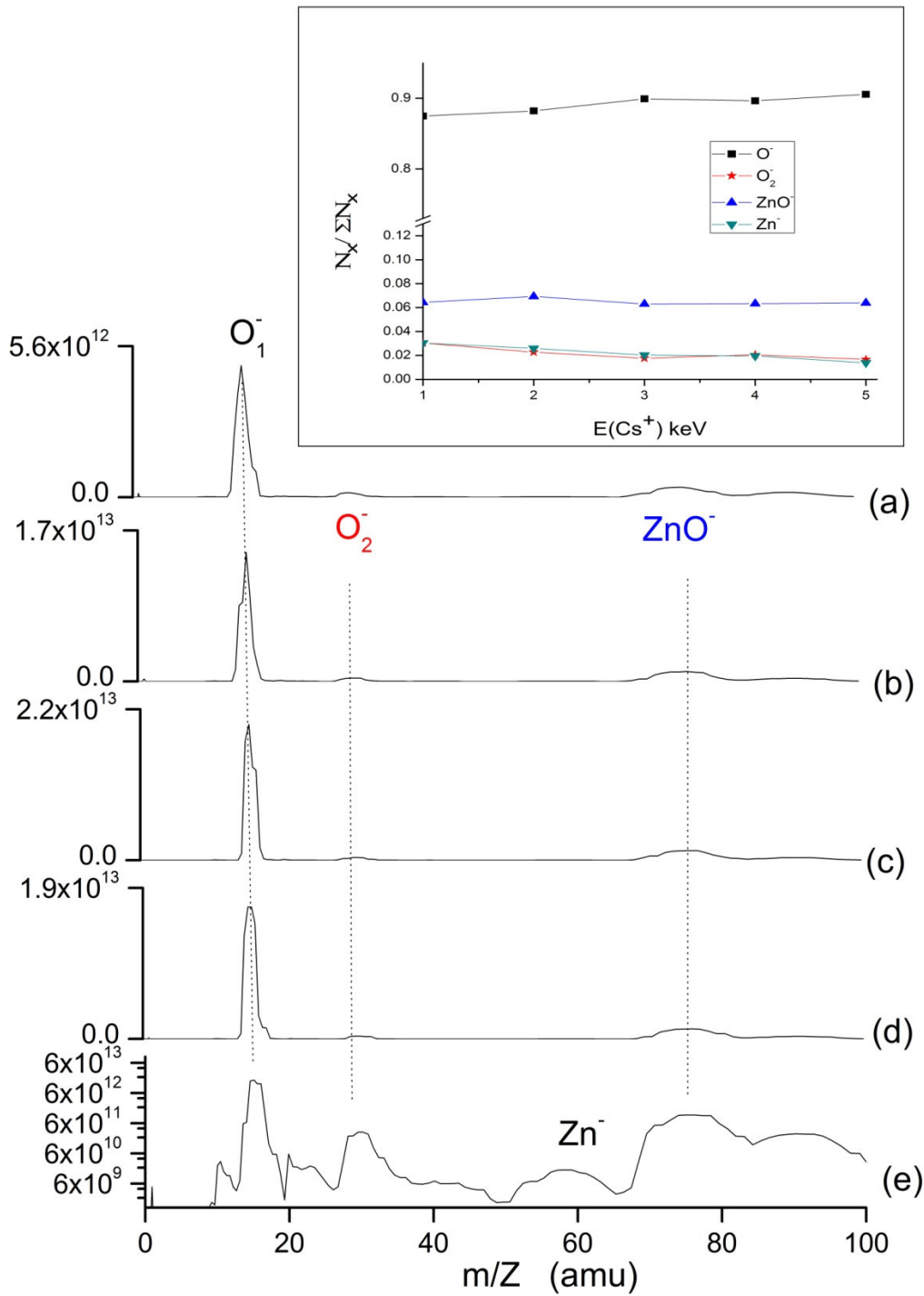

Figure 1. The set of five mass spectra of the sputtered species from Cs$^+$-irradiated ZnO solid is shown (a) to (e) for the increasing cesium energies E(Cs$^+$) from 1.0 keV to 5.0 keV. The atomic O$^-$ has the highest intensity in all the spectra, while Zn$^-$ can be seen only in the last spectra where the anion intensity is shown on log scale. Molecular ZnO$^-$ is the next highest anion after O$^-$. **Inset:** The normalized sputtering yields $N_x/\Sigma N_x$ are plotted as a function of E(Cs$^+$) where $x$ stands for the atomic and molecular species. The normalized yields are independent of Cs$^+$ energy.

We propose the twin mechanisms of emission from the binary collision cascades and thermal spikes to be responsible for the sputtering with atomic and of molecular species from solids like ZnO. Binary collision cascades are initiated with the dissociation of ZnO molecule according to eq. (1). The transition from cascades to thermal spikes can be described by treating the tree of binary cascades as fractal whose dimension increases as the energy of recoils reduces. The fractal dimension can be related to the changes of the effective interatomic potential. Thermal spikes arise as the fractal dimension becomes space-filling[11].

We have indicated elsewhere[12] that the emissions of carbon clusters from fullerenes and carbon nanotubes are most likely the result of ion-induced localized thermal spikes. The mechanism of the origin of spikes is provided in this communication based on the calculated rates of dissociation for the molecular ZnO solid. The equivalence in orders of magnitude between the calculated rates and experimentally measured intensities lend further credence to the existence of spikes and their origins. The 3-D molecular solid of ZnO with interpenetrating Zn and O matrices, is both geometrically and topologically different from the 2-D mono-shelled carbon cages. Therefore, the thermally activated vacancy-generation model may not be directly valid. However, the space-filling character of the cascade-to-spike transition remains intact. In the case of ZnO, the cascades will be the twin-cascades of Zn and O with the interatomic potential $V(r) \propto r^{-\frac{1}{m}}$ where $0 \leq m \leq 1$. The values of m are adjusted to suit the Zn-Zn, O-O and Zn-O collisions. The mean free paths ($\lambda_1, \lambda_2, \ldots \lambda_{n-1}, \lambda_n$) and their ratios are related to the exponent $m$ in the equation $\gamma = \frac{\lambda_n}{\lambda_{n-1}} = 2^{-2m}$ (2). The resulting fractal dimension of the collision cascades is $D = \frac{ln2}{\ln\left(\frac{1}{\gamma}\right)} = \frac{1}{2m}$ (3). These relations show that the cascades' fractal dimension is inversely proportional to the exponent of the interatomic potential. Consequent to the transition of potential V(r) from m=1/2 to 1/4 to 1/6, the fractal dimensions vary from 1 to 2 to 3. When the fractal dimension D and the physical dimension E of the solid are same, the space-filling fractal emerge. For this case the collision mean free paths are of the same order as the interatomic spacing which sets the majority of atoms in motion. This happens in a limited volume where the equipartition of potential and kinetic energies occurs that lead cascades to approach localized thermal equilibrium.

In Figure 2 the rates of the dissociation reaction of ZnO from eq. (1) are plotted as a function of the local spike temperature. The rate equation for the dissociation of ZnO to Zn and O is

$$K = \left\{ [Q(Zn)Q(O)/Q(ZnO)]\exp\left(-\frac{E_{dis}}{kT_{sub}}\right) \right\} \quad (3),$$ where Qs are the respective partition functions per unit volume for the molecule ZnO and its atomic constituents, $E_{dis}$ is the dissociation energy and $T_{sub}$ is the sublimation temperature of ZnO. The value of $E_{dis}$ is taken as 3.5 eV[13] and $T_{sub}$ is expected to be around 2800 K; the calculations show that to be the case. The volume V for the gaseous species has been varied between the limits of a monolayer equivalent to V=$10^{-15}$ m³ to the entire volume of SNICS source V=$10^{-8}$ m³.

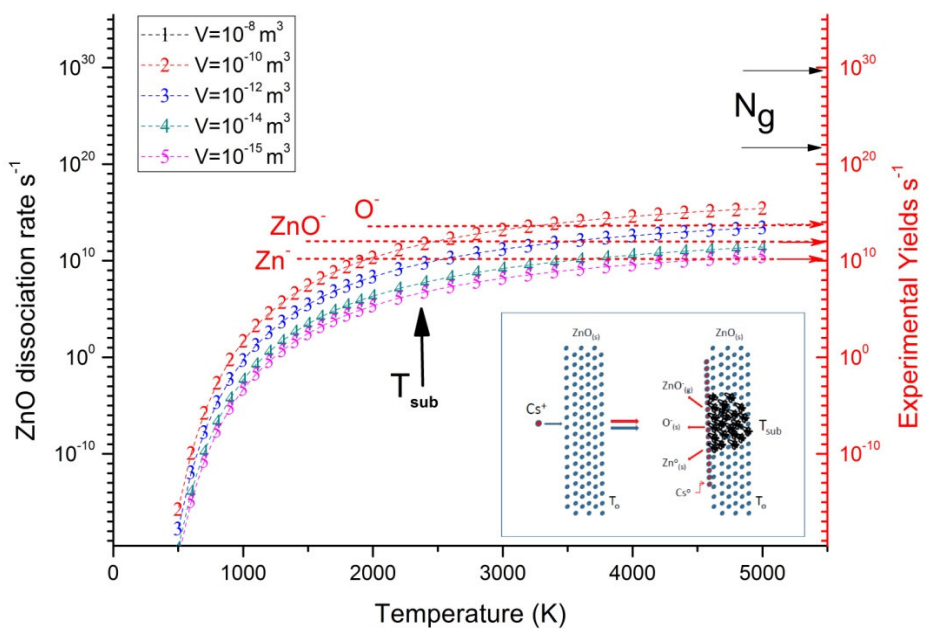

Figure 2. The calculated rates of dissociation of ZnO and the experimental yields of ZnO⁻ are plotted for a range of volume V, as a function of temperature. Experimental yields of O⁻ and Zn⁻ are also indicated. The range of gaseous ZnO from a gas-solid mixture in thermodynamic equilibrium $N_g$, are indicated by arrows. **Inset:** shows the pictorial representation of the emergence of the localized thermal spike.

In Figure 2, in the temperature range from 500 to 1500 K, for the minimum volume V, the ZnO dissociation rates can be seen to vary between $10^{-20}$ to ~$10^2$ s⁻¹. For T >1500 K, the calculated dissociation rates approach the experimentally observed yields of O⁻ and ZnO⁻. With V between $10^{-15}$ and $10^{-12}$ m³, one gets a quantitative agreement between the calculated dissociation rates and the experimentally observed sputtering yields of the atomic and molecular species.

We must make a clear distinction between the localized thermal spike and the whole solid subliming by calculating the number densities required to achieve an overall equilibrium between solid ZnO_s and it gaseous counterpart ZnO_g. When a gas and solid are made up of the same kind of atoms or molecules in a vessel at temperature T, in volume V containing its gaseous species $N_g$. Partition function of the gas and solid $Q_g$ and $Q_s$ can be computed from individual gas molecule ($q_g$) and oscillator ($q_s$)[14]. For the equilibrium condition between the whole solid being in equilibrium with its gaseous components is $N_g = q_g(T,V)/q_s(T)$ (4). This value is indicated for the two extreme values of the volume by two arrows in the top right hand corner of Figure 2. These values of $N_g$ are higher by four to ten orders of magnitude as compared with the experimentally observed number densities of the sputtered species and those calculated from the dissociation rate equation (3) by assuming localized thermal spikes.

We conclude that the sputtering of atomic and molecular species from heavy ion-irradiated molecular solids like ZnO, occurs due to the transition from cascades to thermal spikes. The self-similarity of the branching tree of collisions provides us with the fractal description of this transition that culminates as the space-filling fractals; binary collisions –to- thermal spikes. The space-filling fractal has a local character; hence the thermal spikes are localized around the track of the expanding, energy sharing cascade among the increasing number of neighbors. Each irradiating $Cs^+$ ion in ZnO solid ending in a localized thermal spike while the proximity to the outer surface determining the fate of the subliming species.

Experimental method include sintered ZnO pellets placed in copper bullets that were used as target for NEC's SNICS[15] (Source of Negative Ions by Cesium Sputtering). The ion source is mounted on the 2MV Pelletron at GCU, Lahore. The source was operated with energy of the $Cs^+$ ions $E(Cs^+)$ between 1.0 and 5.0 keV with 1.0 keV steps. Mass spectroscopy was done by an NEC momentum analyzer.

AUTHOR INFORMATION

**Corresponding Author**

*E-mail: sahmad.ncp@gmai.com

**Notes**

The authors declare no competing financial interest.

ACKNOWLEDGEMENTS

The experimental work was performed at the 2 MV Pelletron at CASP, Government College University, Lahore, Pakistan. Technical help provided by Mr. M. Khalil is gratefully acknowledged.